# Topic Level Disambiguation for Weak Queries

**Hui Zhang***

School of Library and
Information Science
Indiana University, U.S.
E-mail: hz3@indiana.edu

**Kiduk Yang**

Department of Library and
Information Science
Kyungpook National University, Korea
E-mail: yangkiduk@gmail.com

**Elin Jacob**

School of Library and
Information Science
Indiana University, U.S.
E-mail: ejacob@indiana.edu

## ABSTRACT

Despite limited success, today's information retrieval (IR) systems are not intelligent or reliable. IR systems return poor search results when users formulate their information needs into incomplete or ambiguous queries (i.e., weak queries). Therefore, one of the main challenges in modern IR research is to provide consistent results across all queries by improving the performance on weak queries. However, existing IR approaches such as query expansion are not overly effective because they make little effort to analyze and exploit the meanings of the queries. Furthermore, word sense disambiguation approaches, which rely on textual context, are ineffective against weak queries that are typically short. Motivated by the demand for a robust IR system that can consistently provide highly accurate results, the proposed study implemented a novel topic detection that leveraged both the language model and structural knowledge of Wikipedia and systematically evaluated the effect of query disambiguation and topic-based retrieval approaches on TREC collections. The results not only confirm the effectiveness of the proposed topic detection and topic-based retrieval approaches but also demonstrate that query disambiguation does not improve IR as expected.

---

**Open Access**

## 1. INTRODUCTION

Information retrieval (IR) has emerged as a central technology in modern society by enabling individuals to extend their ability to discover and obtain knowledge. The quality of queries has a profound impact on retrieval performance because users must formulate their information needs into queries. Queries that produce low retrieval performance on most IR systems are called *weak queries* or *ineffective queries*. In addition to poor query formulation due to the lack of domain knowledge, the problem of weak querying is intensified by the complexities of natural language such as polysemy. For instance, previous research has found that polysemous words in queries can adversely affect automatic query expansion by introducing terms related to incorrect senses of the polysemes (Voorhees, 1994). To address problems of polysemy, IR researchers use techniques developed for word sense disambiguation (WSD) to identify the intended meaning of a given polyseme. However, the effective application of query disambiguation in IR is not a trivial task because the majority of the queries are short and unable to provide the adequate context required by traditional WSD methods.[1] In consequence, previous studies that have examined the issue of query disambiguation report unsatisfactory results (Sanderson, 1994, 2000; Voorhees, 1993).

With the ongoing debate over the benefit of query disambiguation on IR, this study revisits the issue with a new approach that will segment a keyword query into topics and resolve ambiguity only at topic level. The motivation of the study is to examine whether query disambiguation is helpful to IR with the latest developments in machine learning and new knowledge resources such as Wikipedia. In particular, this study focuses on the following research questions,

· Does query disambiguation improve retrieval?
· Do the structural features of a Wikipedia entry (e.g., title, hyperlinks) offer an effective means of establishing context for topic detection and query disambiguation?

with two corresponding hypotheses:

H1. There is no difference in performance between retrieval runs with query disambiguation and baseline retrieval runs without query disambiguation.

H2. There is no difference in performance between retrieval runs with query disambiguation based on Wikipedia knowledge and retrieval runs with query disambiguation based on free text.

The rest of the paper is organized to provide details on the investigation of both questions starting with a brief review of the existing approaches

## 2. LITERATURE REVIEW

Methods for WSD are characterized based on the approach they adopt to acquire word meanings. Approaches that adopt predefined word senses from existing knowledge structures such as dictionaries, thesauri, and Wikipedia are considered to be knowledge-based, while approaches that extract word sense information from the underlying collections are considered to be corpus-based.

Corpus-based WSD approaches derive knowledge from corpora using machine learning algorithms. Researchers who rely on this approach consider word sense disambiguation a straightforward classification problem that attempts to determine the category of a context based on models learned from examples. A Bayesian algorithm has been widely adopted for corpus-based WSD. Gale, Church, & Yarowsky (1992) applied the Naïve Bayesian approach for WSD using a bilingual corpus for training. They selected as the context the 50 words surrounding an ambiguous word. The correct sense was then determined by selecting the sense with the highest probability based on the con-text. The authors found six words in the corpus that had only two senses (e.g., *drug* with the senses of *medical* and *illicit*). The disambiguation model was trained on 60 examples of each of the two senses using words surrounding the target polysemes. The trained model was tested on 90 new word instances for each word sense, and Gale et al. reported an accuracy of 92% on all occurrences of six selected ambiguous nouns. The primary challenge of applying supervised learning algorithms for WSD lies in the overhead for building a sense-tagged corpus. In contrast, knowledge-based approaches overcome this problem by adopting predefined word senses from existing knowledge structures such as dictionaries, thesauri, and lately Wikipedia.

---

[1] The average length of a web query is 2.4 terms and important terms that are descriptive of the information needed are often missing from these short queries (Spink, Wolfram, Jansen, & Saracevic, 2001)

There is a growing trend to use Wikipedia[2] as the sense resource for disambiguation because it provides not only a huge lexicon (i.e., the current English version contains about four million articles) but also extensive descriptions of each word sense. Mihalcea and Csomai (2007) disambiguated terms that appeared in text by mapping them to appropriate Wikipedia articles. The whole process involved two steps: term extraction and word sense disambiguation. In the first step, terms were ranked by their likelihood of becoming hyperlinks in Wikipedia, and only those terms with likelihoods that exceeded a predefined threshold were chosen for disambiguation. For word sense disambiguation, the authors derived word senses from hyperlinks found within Wikipedia articles in order to create training data for supervised disambiguation. For every hyperlink in a Wikipedia article, its author must select the correct destination (i.e., another Wikipedia article) that represents the sense of the anchor text. For example, the term *bar* is linked to different articles on bars in the sense of drinking establishments and in the sense of entertainment sources (e.g., music). Hence, when a polyseme is defined as the anchor of a Wikipedia hyperlink with multiple destinations, a machine learning approach can build a classification model for each polyseme by integrating features such as part of speech and context words. The trained classifier is then used to disambiguate selected terms in the text.

Medelyan, Witten, & Milne (2008) developed another disambiguation approach using Wikipedia as a knowledge resource. They used Mihalcea and Csomai's (2007) strategy of collecting word senses from links in Wikipedia articles. To disambiguate a polyseme in the text, the surrounding terms that were unambiguous Wikipedia anchors (i.e., links to only one Wikipedia article) were then chosen as contexts. The disambiguation process was carried out by calculating semantic similarity to the contexts and conditional probability for the polyseme. Semantic similarity was calculated as the average of each candidate article that represented a sense of the polyseme to all context articles. Conditional probability was computed from counts of Wikipedia links; for example, because the word *jaguar* links to the article for the car jaguar in 466 out of 927 links, the probability of the automobile sense is 0.5. Semantic disambiguation was then determined by multiplying the semantic similarity by the conditional probability of each sense and selecting the one that produced the highest score. The approach devised by Medelyan et al. achieved an *F*-measure of 0.93 on automatically mapping terms to correct Wikipedia articles.

The latest development in query disambiguation research emphasizes the integration of natural language technology and knowledge bases. Selvaretnam and Belkhatir (2012) proposed a query expansion framework considering sense disambiguation as an essential step. In particular, the authors suggested that syntactic parsing such as part of speech recognition should be applied at first followed by semantic sense disambiguation. The decision of whether a sense is appropriate will depend on the relatedness of that sense to its occurring context measured in an external knowledge base such as WordNet. Similar to this study, Klyuev & Haralambous (2011) investigated query disambiguation in the context of improving query expansion, with their finding indicating that combining multiple knowledge bases such as WordNet and Wikipedia could bring better retrieval results. The authors also suggested that selecting the Wikipedia articles that are closely related to the query is vital to the success of query expansion, which affirms the importance of query disambiguation if one query term has multiple matched Wikipedia articles.

---

[2] http://www.wikipedia.org/

Lack of context deepens the ambiguity problem for weak queries. Most current disambiguation approaches demand 20 to 50 context words to produce a relatively reliable prediction. However, weak queries are unable to provide that much context because they usually contain only two or three words. A common solution is to enrich a short query with query expansion techniques such as pseudo relevance feedback and web expansion. The challenge when using these strategies is how to overcome the impact of "noisy" words due to inaccurate query expansion. Both query expansion and query disambiguation can benefit from operating at the level of the topic instead of the word. Query expansion at the topic level will reduce error due to query drift and lead to higher retrieval performance (Bendersky, Croft, & Smith, 2009). One major problem for query disambiguation is how to find the appropriate context for a polyseme. Using query topic detection, it is likely that a polyseme and its most revealing context will be grouped in the same segment based on co-occurrence patterns, and this will help a software agent to improve its disambiguation accuracy (Navigli, 2009). Recognizing boundaries and the correct meanings of query topics is an important step towards understanding a user's search intent and improving the retrieval accuracy of weak queries. The unsupervised approach is efficient, but it has a high error rate. One solution that could lower the error rate would be to use external knowledge for guidance. The research reported here explored this direction by developing an approach to unsupervised query segmentation that utilized the knowledge in Wikipedia to achieve better performance on boundary recognition.

## 3. METHODOLOGIES

### 3.1. Topic Detection with Language Model

The first challenge for topic-based query disambiguation is to develop an approach to recognize top-ics, such as phrases and named entities, from user queries. Given the success of the statistical language model (LM) with tasks involving natural language, including IR (C. X. Zhai, 2008), this study developed an LM-based approach as a solution for the challenge.

Each keyword query will be split into complementary pairs of n-grams, or "chunks," as input for the topic detection process. For instance, the query *new york* and *times square* will be segmented into pairs of chunks as: *[new][york times square], [new york][times square]*, and *[new york times][square]*. This method is effective on weak queries where typically there are fewer topics in each query in consequence of the short query length.

The topic detection algorithm first expands the query with textual contents from the Web to address the problem of lack of context, which is common in user queries. In particular, a Google search can be performed using the original query,[3] and the textual content extracted from the first 20 results of the search can be used to provide context for the query. With this enriched context $C$ and a set of topic candidates $S$, the question of topic detection is stated as: Given a query $q$, which candidate $s$ in $S$ generates the highest probability? This question can be formulated as the following:

$$p(s|q) = \frac{p(s,q)}{p(q)} \qquad (1)$$

Because $q$ is random, $p(s|q)$ is only determined by $p(s,q)$. To estimate the value $p(s,q)$, the assumption is that they are drawn from the context collection $C$:

$$p(s,q|C) = \sum_{d \in C} p(s,q|d).p(d|C) \qquad (2)$$

where $p(d)$ is a prior distribution of documents and is assumed to be uniform across all documents, so the major challenge left is how to estimate the probability of $p(s,q|d)$:

$$p(s,q|d) = p(s|q,d).p(q|d) \qquad (3)$$

where $p(q|d)$ is estimated with maximum likelihood

---

[3] https://www.google.com/

estimation as:

$$p(q|d) = \prod_{w \in q} p(w|d) \qquad (4)$$

A serious problem imposed on LM-based approaches is data scarcity, which is worsened by the textual content because a few words occur frequently while many words appear rarely or are entirely absent in the document collection. Jelinek-Mercer smoothing, which has been proven to be effective for IR tasks (C. Zhai & Lafferty, 2004), is chosen as the method to estimate a "discount" probability to words that are not present. With Jeliner-Mercer smoothing, $p(s|q,d)$ is transformed into:

$$p(s|q,d) = (1 - \alpha_D)p(s|q,d) + \alpha_D p(s) \qquad (5)$$

where $p(s|q,d)$ is estimated as:

$$p(s|q,d) = \frac{f(s,q,d)}{\sum_{w \in q} f(w,d)} \qquad (6)$$

and the background model $P(S)$ is estimated as with Wikipedia titles and anchors:

$$P(S) = \lambda P(S|Wk\_title) + (1-\lambda)P(S|Wk\_anchor)$$
$$= \lambda \frac{f_{S,title}}{|title|} + (1-\lambda) \frac{f_{S,anchor}}{|anchor|} \qquad (7)$$

In summary, the generative probability of a query topic $s$ that appears in query $q$ given the context of collection $C$ is estimated as

$$p(s,q|C) = \sum_{d \in C} [(1 - \alpha_D)p(s|q,d) + \alpha_D p(s)]p(q|d) \qquad (8)$$

and the candidate (i.e., query chunk) with the highest probability will be chosen as the query topic.

## 3.2. Topic Disambiguation

The first task of word sense disambiguation is to build what is known as a "sense inventory" containing all the possible meanings for each polyseme. Follow-ing the strategy of Mihalcea and Csomai (2007), the approach adopted here collects word senses as hyperlinks (i.e., anchors) in Wikipedia articles. For many hyperlinks in Wikipedia the author manually annotates the intended meaning of an anchor by linking it to a relevant Wikipedia article. Therefore, the sense inventory for any polyseme can be derived by extracting link destinations from all hyperlinks associated with the polyseme. Using this approach, five senses were identified for the polyseme *bar: bar (counter), bar (establishment), bar (landform), bar (law),* and *bar (music).*

The disambiguation approach used in this research is known as a *decision list* (DL), which contains a set of ordered and conjunctive rules that are either handcrafted or derived by algorithm. For each ambiguous query topic, the algorithm first uses the count of overlapping Wikipedia anchors between the topic and the rest of the query terms to resolve its appropriate meaning. If no overlapping anchor exists, the algorithm will consider the count of co-occurrence words in the definition paragraph (excluding stopwords) between the topic and the other query terms.[4] And the fallback criterion is the frequency of different senses appearing in Wikipedia, where the most common sense is chosen as the disambiguation result. The disambiguation algorithm is illustrated in Figure 1 with details.

## 3.3. Topic-based Query Expansion

One major problem related to weak queries is the failure to respond to required aspects of a user's information need (Buckley & Harman, 2004); but adding inappropriate expansion terms to a weak query can lead to the problem of query drift and may reduce the performance of query expansion (Ruthven & Lalmas, 2003). To address problems of query drift, this research introduces a query expansion method that would provide for robust retrieval by exploiting the query topics detected in the disambiguation process.

The query expansion method used here relied on terms identified both in Wikipedia and on the Web in order to harvest evidence from two different types of content. Wikipedia is considered a structured resource because the full text of each article was excluded from the process of query expansion: Given that expansion terms were extracted only from the title of a Wikipedia article and its definition paragraph (i.e., the first paragraph of a Wikipedia article) based on the frequency, they are not only concise and accurate but also provide a relatively small vocabulary. In contrast, even though expansion terms extracted from the Web would include terms that were potentially irrelevant, the fact that they are harvested from a large vocabulary and can therefore address aspects of the topic missing in Wikipedia articles was considered an advantage. Thus, this approach is able to exploit the advantages of both sources to offset the weakness of each. The technique chosen for Web-based query expansion is local context analysis (LCA), which selects concepts based on their co-occurrence with query terms and their frequency in the whole collection.

---

[4] The first paragraph of a Wikipedia article that is mandatory in order to provide a brief summary of the subject.

1. Build sense inventory: Search the Wikipedia knowledge base to identify all articles (destinations) linked to by the query segment as anchor text and count the total number of each unique destination article; save the count; a query segment is defined as a polyseme if it has more than one unique destination article.

2. Build sense representation:
   a. For sense as page link: For each linked-to page, extract and save all the Wikipedia anchors that appear in the article.
   b. For sense as definition: Extract and save the text of the definition paragraph of each linked-to page.
   c. For sense as count: count the occurrences of every unique linked-to page in Wikipedia.

3. Disambiguation:
   **For** each query segment
   **if** it is a polyseme
   · get the senses built in step 2
   · search Wikipedia to find whether the query context matches any article title or anchor text
   · count the number of overlapping linked-to articles between the Wikipedia pages of the query context and each sense of the query segment
      **if** there are overlapping links
         · identify the sense of the query segment as the Wikipedia page that has the highest overlapping count with the query context page;
   **else**
         · identify the sense of the query segment as that of the Wikipedia page whose definition paragraph has the highest number of co-occurring words with the query context page;
      **if** there are no overlapping words
         · identify the sense of the query segment as the one that appears most frequently in Wikipedia based on the results of step 1.

**Fig. 1** Disambiguation Algorithm for Query Segment

# 4. EXPERIMENT DESIGN

## 4.1. Data

The test collections used in the experiment are AQUAINT and Blog06. The AQUAINT corpus is distributed by the Linguistic Data Consortium (LDC) and has been used in TREC competitions as the test collection for both the HARD Track and the Robust Track in 2005.[5] The AQUAINT corpus consists of 1,033,461 news stories taken from the *New York Times*, the Associated Press, and the Xinhua News Agency newswires between the years 1996 and 2000. The corpus contains

---

a total of 284,597,335 terms, 707,778 of which are unique, and the mean document length is 275 words (Baillie, Azzopardi, & Crestani, 2006). The Blog06 collection was developed by the University of Glasgow and consists of 148GB of blog data spanning a time period of eleven weeks from December 2005 to February 2006. The Blog06 collection contains more than 3.2 million permalinks[6] that the Blog Track guideline[7] considers eligible retrieval units and was used as the test collection for the TREC Blog Track in 2006, 2007, and 2008. Both collections were indexed and searched by Indri,[8] a search engine that combines the advantages of a language model and an inference network that allows the formulation of structured queries composed by topics (Metzler & Croft, 2004).

This research used *mean average precision* (MAP) as the measure of IR effectiveness. *Average precision* is a single-valued measure used to evaluate a system's overall performance for a given query. It is calculated by dividing the sum of the precision values obtained after retrieval of each relevant document by the total number of relevant documents in the collection. Because average precision rewards an IR system that ranks a relevant document higher, it is compatible with the overall goal of this research, which is to improve retrieval accuracy. System performance over a set of queries can be evaluated using MAP-the mean value of average precision values over all queries. In addition, pairwise differences between systems can be plotted for each query to indicate relative system performance over queries.

## 4.2. Retrieval with Query Expansion

The process of query expansion used the detected query topics as input and produced as output a list of weighted expansion terms for each query. The expansion terms were based on the detected query topics and were selected from both the Web and Wikipedia. The final version of the transformed query was formulated according to the Indri format and included three components: the original query, the detected query topics, and the top 20 expansion terms as determined by the weighting. The query topic component was constructed according to the following rules: If topics were detected in the query, then each topic would be formulated as an exact match, and all words in the original query would be included in an unordered match within a window of 12 words (i.e., *uw12*); if no topic were detected, then the words in the original query would be formulated as an unordered match within 20 words (i.e., *uw20*). For instance, given the query *law enforcement dogs* where *law enforcement* is the topic, the query would be transformed into *#1(law enforcement)* and *#uw12(law enforcement dogs)*. In contrast, the query *marine vegetation* would be transformed into *#uw20 (marine vegetation)* because the two words do not constitute a topic. Each component of the final query was manually assigned a weight to quantify its contribution to relevance judgment, and the weights for the original query, the query topic, and the expansion terms were specified as 0.5, 0.2, and 0.3, respectively, based on preliminary results. Figure 2 illustrates the final version of the query *law enforcement dogs*.

## 4.3. Query Disambiguation Experiment

The first of these experiments (referred to hereafter as experiment #1) investigated the effect of query disambiguation on IR by comparing the retrieval performance of ambiguous queries with and without resolving topic

---

```
#weight(0.5 #combine(law enforcement dogs)
        0.2 #combine(#1(law enforcement) #uw12(law enforcement dogs))
        0.3 #weight(0.170731707317073 dog 0.146341463414634 police 0.0487804878048781 morphology ...))
```

**Fig. 2** Example of a fully transformed query using Indri syntax

---

ambiguity (i.e., H1). Four ambiguous queries from the Blog collection and ten ambiguous queries from the HARD collection were selected based on the criterion that either the query or, at minimum, one topic in the query was a Wikipedia hyperlink referring to one or more Wikipedia articles. For example, the string "black bear" could refer to a minor league baseball team or to a North American animal species. Query disambiguation effects on IR for ambiguous queries from the Blog and HARD collections were analyzed using four different query treatments: *wsd_qe* represents query expansion (*qe*) with disambiguation (*wsd*) based on Wikipedia hyperlinks where the resolved Wikipedia article is used as the source for query expansion terms; *no_wsd_qe* represents query expansion without disambiguation and uses Google search results as the source for query expansion terms; *wsd _ir* represents information retrieval (*ir*) with query disambiguation and formulates the query with the disambiguated segment(s) as the topic;[9] and *no_wsd_ir* represents word-based retrieval without disambiguation.

## 4.4. Wikipedia for Query Disambiguation Experiment

The other experiment (referred to hereafter as experiment #2) investigated the hypothesis that using Wikipedia structures for query disambiguation would lead to better IR performance (i.e., H2). IR performance was measured by comparing the retrieval performance produced with two disambiguation approaches: results obtained by using Wikipedia knowledge and results obtained by using only textual context. Experiment #3 used only ambiguous query topics that had more than one sense representation derived from Wikipedia. For each ambiguous topic, this experiment attempted to identify the correct sense using two types of contexts: Wikipedia hyperlinks and free text (i.e., terms appearing in the first paragraph of the associated Wikipedia article). The assumption was that the two types of contexts would lead to significantly different disam-biguation results (i.e., would not resolve to the same Wikipedia article). If a Wikipedia article was found, the polyseme under consideration would be considered a valid topic and the Wikipedia article would become the source of query expansion terms; if a Wikipedia article was not found, the polyseme under consideration would be treated as text without query expansion.

Queries from the Blog and HARD collections were disambiguated using knowledge at two different semantic levels: Wikipedia and free text. These experiments used four different query treatments:

· *wsd_qe_wiki*[10] represents query disambiguation (*wsd*) with Wikipedia knowledge (*wiki*) such as hyperlinks in the articles using the Wikipedia article that expresses the correct sense as the source for query expansion (*qe*) terms; if no associated Wikipedia article was found, the original query was used for retrieval without expansion;

· *wsd_qe_nowiki* represents query disambiguation with text context (i.e., expansion terms acquired by Local Context Analysis from Google search results, or *nowiki*) to identify word sense using the Wikipedia article that expresses the correct sense as the source for query expansion; if no associated Wikipedia article was found, the original query was used for retrieval without expansion;

· *wsd_ir_wiki* represents query disambiguation using Wikipedia knowledge such as hyperlinks in the articles to formulate the disambiguated query segment as a topic for information retrieval (*ir*); if no associated Wikipedia article was found, the query segment in consideration was treated as text;

· *wsd_ir_nowiki* represents query disambiguation using text context (i.e., the rest of the query) to identify word sense and formulate the disambiguated query segment as a topic for information retrieval; if no associated Wikipedia article was found, the query segment was treated as text.

---

[9] For instance, the disambiguated segment *Whole Foods* in the query *Whole Foods wind energy* will be formulated as a topic in Indri query *#combine(#1(Whole Foods) #1(wind energy))*.

[10] The experiment *wsd_qe_wiki* follows the same procedure as outlined in *wsd_qe* in section 4.3; the additional suffix of *_wiki* is used to highlight the source of knowledge, which is the focus of the experiment. The same naming patterns also apply to *wsd_ir_wiki*.

Retrieval performance was measured by MAP in both experiments.

# 5. RESULTS

## 5.1. Query Disambiguation to IR Improvement

A summary of the results for retrieval with and without query disambiguation are provided in Table 1. The Wilcoxon signed ranks test performed on the results did not reject the null hypothesis (H1) that query disambiguation has no significant effect on retrieval performance for either the Blog queries or the HARD queries. For ambiguous queries from the Blog collection, Wilcoxon tests for two-tailed significance conducted between *wsd_qe* and *no_wsd_qe* and between *wsd_ir* and *no_wsd_ir* retained the null hypothesis ($p$=0.068, $N$=4) even though, in all experimental runs, disambiguated queries outperformed queries that had not been disambiguated. For ambiguous queries from the HARD collection, Wilcoxon tests for two-tailed significance conducted between *wsd_qe* and *no_wsd_qe* and between *wsd_qe* and *no_wsd_ir* also retained the null hypothesis, but at a higher p-value ($p$>0.2, $N$=10).

## 5.2. Wikipedia Effect on Query Disambiguation

Retrieval results for queries from the Blog and HARD collections are listed in Table 2 when disambiguation is carried out using Wikipedia knowledge or free text. For both the Blog and the HARD queries, Wilcoxon's signed ranks test did not reject the null hypothesis (H2) that using Wikipedia knowledge had little effect on query disambiguation. Comparison of the *wsd_qe_wiki* and *wsd_qe_nowiki* treatments and the *wsd_qe_wiki* and *wsd_ir_nowiki* treatments for Blog queries using Wilcoxon's signed ranks test retained the null hypothesis (0.2>$p$>0.1) with a sample of four values ($N$=4); comparison of *wsd_qe_wiki* and *wsd_qe_nowiki* treatments and of the *wsd_qe_wiki* and *wsd_ir_nowiki* treatments for the HARD queries using Wilcoxon's test retained the null hypothesis (0.8>p>0.1) with a sample of ten values (N=10).

# 6. DISCUSSION

Experimental results indicate that both of the null hypotheses regarding query disambiguation should be retained: Query disambiguation has little impact on IR; and the use of knowledge in Wikipedia documents does not produce significant improvement in disambiguation accuracy. Furthermore, this result is found for queries from both the Blog and the HARD collections.

**Table 1.** Query disambiguation effects on IR for ambiguous queries from Blog (denoted B) and HARD (denoted H) collections

| | Mean Average Precision | | | |
| | *wsd_qe* | *no_wsd_qe* | *wsd_topic_ir* | *no_wsd_ir* |
|---|---|---|---|---|
| Fox News Report (B) | 0.1363 | 0.1253 | 0.1397 | 0.1253 |
| Business Intelligence Resources (B) | 0.0671 | 0.0464 | 0.0660 | 0.0437 |
| Whole Foods wind energy (B) | 0.7927 | 0.6553 | 0.7928 | 0.7044 |
| federal shield law (B) | 0.6125 | 0.2514 | 0.5759 | 0.1167 |
| Radio Waves and Brain Cancer (H) | 0.2854 | 0.3028 | 0.0992 | 0.0982 |
| Black Bear Attacks (H) | 0.5066 | 0.5142 | 0.5504 | 0.5181 |
| mental illness drugs (H) | 0.0801 | 0.0790 | 0.0496 | 0.0594 |
| Ireland peace talks (H) | 0.2318 | 0.2431 | 0.2518 | 0.2568 |
| Legal Pan Am, 103 (H) | 0.3830 | 0.3145 | 0.2795 | 0.3682 |
| law enforcement dogs (H) | 0.1242 | 0.0837 | 0.1132 | 0.0968 |
| Greek philosophy stoicism (H) | 0.4403 | 0.5220 | 0.5202 | 0.5271 |
| Inventions scientific discoveries (H) | 0.1173 | 0.1099 | 0.0884 | 0.0966 |
| family leave law (H) | 0.6549 | 0.6702 | 0.5902 | 0.5030 |
| tax evasion indicted (H) | 0.1197 | 0.1277 | 0.053 | 0.1045 |

**Table 2.** Effects of using Wikipedia knowledge for disambiguation of queries from Blog (denoted B) and HARD (denoted H) collections

| | Mean Average Precision | | | |
|---|---|---|---|---|
| | wsd_qe_wiki | wsd_qe_nowiki | wsd_ir_wiki | wsd_ir_nowiki |
| **Fox News Report (B)** | 0.1339 | 0.1253 | 0.1397 | 0.1253 |
| **Business Intelligence Resources (B)** | 0.0558 | 0.0558 | 0.0660 | 0.0660 |
| **Whole Foods wind energy (B)** | 0.7927 | 0.7919 | 0.7928 | 0.7928 |
| **federal shield law (B)** | 0.6366 | 0.1167 | 0.5759 | 0.1167 |
| **Radio Waves and Brain Cancer (H)** | 0.0952 | 0.0952 | 0.0992 | 0.0982 |
| **Black Bear Attacks (H)** | 0.5042 | 0.5181 | 0.5504 | 0.5181 |
| **mental illness drugs (H)** | 0.0791 | 0.0791 | 0.0496 | 0.0496 |
| **Ireland, peace talks (H)** | 0.2138 | 0.2568 | 0.2518 | 0.2568 |
| **legal, Pan Am, 103 (H)** | 0.383 | 0.3682 | 0.2795 | 0.3682 |
| **law enforcement, dogs (H)** | 0.1242 | 0.1058 | 0.1132 | 0.1132 |
| **Greek, philosophy, stoicism (H)** | 0.4488 | 0.4050 | 0.5202 | 0.5202 |
| **inventions, scientific discoveries (H)** | 0.1036 | 0.1036 | 0.0884 | 0.0884 |
| **family leave law (H)** | 0.6922 | 0.6922 | 0.5902 | 0.5902 |
| **tax evasion indicted (H)** | 0.1197 | 0.1197 | 0.053 | 0.053 |

One of the major motivations for this research is to re-examine the ongoing argument regarding the usefulness of query disambiguation in IR in light of relatively new knowledge resources such as Wikipedia. As has been pointed out in the IR literature (Harman, 1992; Salton & Buckley, 1990), disambiguation accuracy and quality of the sense inventory have been thought to make major contributions to disambiguation effects on IR. Given the experimental results indicating that the null hypotheses regarding query disambiguation should be retained for both query collections, it is helpful to examine the influence of each of these two factors in the context of the current findings.

### 6.1. Correlation between Disambiguation Accuracy and IR Performance

Query disambiguation accuracy has been claimed to be the most important factor affecting the effectiveness of disambiguation in IR (Buckley, Salton, Allan, & Singhal, 1995; Salton & Buckley, 1990). Based on the commonly held assumption that low accuracy in the resolution of polysemes will hurt retrieval performance, an experiment was designed to examine whether there was a correlation between disambiguation accuracy and retrieval performance: For an ambiguous query term, each of its word senses (i.e., a corresponding Wikipedia article) was represented by three features: anchor links that appeared in a Wikipedia article; the text in the first paragraph of an article; and the count of each sense appearing in Wikipedia. The hypothesis was that there would be discrepancies in disambiguation accuracy across the different sense representations, and the goal was to test whether retrieval performance would be significantly affected by ineffective query disambiguation.

Only one query from the Blog collection (i.e., _Whole Foods wind energy_) and two queries from the HARD collection (i.e., _Ireland, peace talks_, and _Greek, philosophy, stoicism_) yielded different disambiguation results. A list of retrieved documents was produced for the disambiguated retrieval results for each of these three queries using the corresponding Wikipedia article(s) as the source of query expansion terms. Assuming that only one sense represented by the correct Wikipedia page was appropriate given the query context, it was possible to observe whether disambiguation errors had an impact on retrieval, as shown in Table 3.

**Table 3.** Retrieval performance of disambiguation accuracy measured as MAP compared to baseline. Wikipedia page ids are indicated in parentheses and correct Wikipedia page ids are presented in bold

| | Mean Average Precision | | | |
|---|---|---|---|---|
| | baseline | represent_by_link | represent_by_text | represent_by_count |
| **Ireland, peace talks (HRAD)** | 0.2568 | 0.2318 (**3021179**) | null | 0.2063 (7258068) |
| **Greek, philosophy, stoicism (HARD)** | 0.5271 | 0.4403 (**10649725**) | 0.4403 (10649725) | 0.4002 (171171) |
| **Whole Foods wind energy (Blog)** | 0.7044 | 0.7927 (**620343**) | 0.7927 (620343) | 0.7002 (30871513) |

To examine the impact of query disambiguation accuracy on retrieval performance, three statistical significance tests (i.e., t-Test with paired samples) were carried out on the results listed in Table 3:[11] baseline vs. query expansion based on the correctly disambiguated Wikipedia page, which is in bold; baseline vs. query expansion based on the incorrectly disambiguated Wikipedia page; and query expansion based on the correctly disambiguated Wikipedia page vs. query expansion based on the incorrectly disambiguated Wikipedia page. The resulting *t* values for all three tests were smaller than the critical values for the *0.05* significance level, and the null hypothesis was retained. Although the results of the t-Test with paired samples indicate that disambiguation accuracy does not have a significant impact on retrieval performance, the null hypothesis should not be rejected out of hand due to the small sample size. For instance, the term *Whole Foods* is the title of several Wikipedia pages; but, given the query context of *wind energy*, only the Wikipedia article with the page id *620343* is relevant and leads to better retrieval performance when compared to the baseline retrieval results (i.e., query expansion run without disambiguation). In addition, a mistake in query disambiguation such as pointing to Wikipedia page id *30871513* for the query *Whole Foods wind energy* produced a decrease in retrieval performance when compared to retrieval performance using the appropriate Wikipedia page for disambiguation.

## 6.2. Wikipedia Effect for Query Disambiguation

Any attempt at resolving natural language ambiguities will depend on the quality and scale of the sense inventory, the repository of terms, and the common meanings for each term. WordNet is an example of a sense inventory and has been used extensively for word sense disambiguation (Jing & Croft, 1994; Qiu & Frei, 1993). However, according to recent studies (Rada Mihalcea, 2003; Prakash, Jurafsky, & Ng, 2007), WordNet has certain major drawbacks that make it unsuitable for query disambiguation in IR:

· Sense granularity is too fine: For purposes of logic reasoning, definitions in WordNet include very fine distinctions between word senses. For instance, the verb *eat* has the two senses *take in solid food* and *eat a meal*. While this fine sense granularity is obviously helpful in areas such as artificial intelligence, it is not necessary in IR.

· Semantic connections are too complete: The number of relationships defined in WordNet produces a huge number of possible semantic connections between two words. Such a large number of semantic connections will overload the IR system, requiring longer processing time without increasing relevance.

· The scale of WordNet is too limited: The latest version of WordNet (i.e., WordNet 3.0) contains a total of 155,287 words and 206,941 word-sense pairs.[12] Such a limited scope is not adequate for modern IR applications.

---

[11] If two sense representations yield the same disambiguation results, such as page id 10649725 for query *Greek philosophy stoicism*, only one result is highlighted in bold.
[12] http://wordnet.princeton.edu/wordnet/man/wnstats.7WN.html

Given the weaknesses of WordNet, researchers have been looking for other knowledge resources that could be used for query disambiguation. Because of its large scale and the richness of its content, Wikipedia is a primary candidate and was selected for this research motivated by the hypothesis that the coverage and currency of Wikipedia articles would improve query disambiguation accuracy. However, as indicated by the results presented in Table 2, it is evident that using the structural knowledge in Wikipedia (i.e., the anchor links) did not lead to significant improvement in disambiguation accuracy because user queries do not normally contain sufficient context to associate a query with an appropriate Wikipedia page.

To understand the failure of Wikipedia as a knowledge base for disambiguation, it is helpful to assess Wikipedia based on the same criteria that have been applied to WordNet:

· Sense granularity: Word senses are indicated by article topics in Wikipedia. For instance, in Wikipedia, the word *Bush* has various senses ranging from a type of plant, to a surname, and even to an island because there is a Wikipedia article for each of these three senses as discrete topics. Each sense should have adequate context for disambiguation in natural language, which is a decided advantage of Wikipedia. However, sense granularity is still an important issue in Wikipedia because there is no oversight of or planning for the nomination and selection of topics. Anyone can add a new meaning for a word by contributing a new Wikipedia article for that topic; the only constraint on a new page is that it must follow Wikipedia's guidelines for articles. In consequence, a disambiguation program could spend unnecessary processing time on senses that are rarely used or incorrectly identify a sense because terms expressed in the query are either undefined or over-defined in Wikipedia.
· Semantic connections: In Wikipedia, polysemes are generally connected to meanings through one of two methods: through anchor links that point to an article associated with one sense of the term or through a so-called *disambiguation page* that lists all meanings associated with a term. Harvesting word senses from anchor links has two advantages over a disambiguation page: It associates a meaning within the language context where it occurs, and it offers a count distribution of sense usage in Wikipedia. Experimental results show that distribution of sense usage is effective in query disambiguation since the primary sense (i.e., the sense that appears most frequently) is usually the correct one. However, no matter which sense representation is used, query disambiguation will suffer from the weaknesses of sense granularity.
· Scale: As of December 2012, the English version of Wikipedia contained more than four million articles, with new articles submitted every day. However, it requires significant effort both to extract the knowledge embedded in Wikipedia articles and to build that knowledge into a structural resource for applications such as query disambiguation.

Analysis of the problems associated with granularity, sense connections, and scale in both WordNet and Wikipedia indicate that query disambiguation demands a type of knowledge resource that is very different from what either of these resources offer. For instance, based on the observation that users will correct spelling errors, add context, or change words in original queries to improve retrieval results (Guo, Xu, Li, & Cheng, 2008), large-scale query logs available from commercial search engines could be used to extract a series of queries from individual sessions and build a knowledge base that would not only catch grammatical variations and misspellings but also semantic contexts such as synonyms and co-occurring terms that point to the same topic (i.e., the same document). It would also be possible to construct a sense inventory by harvesting click-through records from query logs. Such a knowledge base built from query logs would not only save the cost of creating definitions and samples manually but might also be more effective for query disambiguation. In fact, using well-developed data mining algorithms, a knowledge base generated from a large and diversified query log, would be a very special kind of mass intelligence pro-duced by user collaboration on a common task.

## 7. CONCLUSION

To overcome the challenge of ineffective user queries, this research implemented a query disambiguation approach that integrates topic detection and maps the detected topic to the most appropriate Wikipedia page. This research tested two hypotheses for query disambiguation in IR. The experimental results could not reject the null hypothesis that there was no significant difference in performance between retrieval with query disambiguation and retrieval without disambiguation. Furthermore, statistical testing did not support the hypothesis that representing word meanings with structural Wikipedia knowledge such as anchor links would significantly improve disambiguation effectiveness in IR compared to representing meanings with text. Both of these findings suggest that future knowledge bases of word meanings should favor defining word senses by harvesting language usage patterns, probably from large search engine logs, in order to maintain a rich level of diversified contexts for query disambiguation and optimization. However, because the test collections used in this research contained only a limited number of ambiguous queries,[13] the validity of any conclusions regarding query disambiguation based on statistical analysis must be considered preliminary due to the small sample size.

---

[13] These are four ambiguous queries in the Blog collection and ten queries in the HARD collection.